\documentclass[12pt,twoside]{article}
\usepackage[utf8]{inputenc}
\usepackage[english]{babel}
\usepackage{hyperref}
\usepackage{amsthm,thmtools}
\usepackage{amsfonts}
\usepackage{amssymb}
\usepackage{amsmath}
\usepackage{float}
\DeclareMathOperator\arctanh{tanh^{-1}}
\newcounter{enunciato}[section]
\newtheorem{ittheorem}{Theorem}

\newtheorem{itlemma}{Lemma}

\usepackage[figurename=Fig]{caption}
\usepackage[labelfont=bf]{caption}
\usepackage[labelsep=period]{caption}
\usepackage{caption}
\usepackage{subcaption}
\usepackage{graphicx}
\usepackage{geometry}
\usepackage{subfiles}
\usepackage{float}
\usepackage{mathtools}
\usepackage{pgfplots}
\pgfplotsset{/pgf/number format/use comma,compat=newest}
\usepackage{url}
\usepackage{tikz}
\usepackage{authblk}
\usetikzlibrary{arrows,decorations.pathmorphing,backgrounds,positioning,fit,petri,patterns}
\title{Human-AI ecosystem with abrupt changes as a function of the composition}
\author[1]{Pierluigi Contucci}
\author[2]{János Kertész}
\author[1]{Godwin Osabutey}
\affil[1]{Department of Mathematics, University of Bologna}
\affil[2]{Department of Network and Data Science, Central European University}
\date{\today}

\begin{document}

\maketitle

\begin{abstract}
The progressive advent of artificial intelligence machines may represent both an opportunity or a threat. In order to have an idea of what is coming we propose a model that simulate a Human-AI ecosystem. In particular we consider systems where agents present biases, peer-to-peer interactions and also three body interactions that are crucial and describe two humans interacting with an artificial agent and two artificial intelligence agents interacting with a human. We focus our analysis by exploring how the relative fraction of artificial intelligence agents affect that ecosystem. We find evidence that for suitable values of the interaction parameters, arbitrarily small changes in such percentage may trigger dramatic changes for the system that can be either in one of the two polarised states or in an undecided state.

\end{abstract}

\section{Introduction}
Artificial Intelligence (AI) is the property of human-made systems with specific or general goals and the ability to perceive and process information from their environment, to take action towards achieving their goals. Fields of applications include understanding human speech, pattern recognition, recommendation systems, navigation, health care, military and many more~\cite{Russel2021AI}. Such systems have become ubiquitous during the recent decades and are influencing our lives in many ways. Consciously or not, human-AI interactions have become natural in several domains like e-commerce, medical diagnostics, self-driven vehicles or online social networks. 

The economic, sociological and ethical aspects of this emergent, complex human-AI ecosystem are of eminent interest, yet, we are only at the beginning of their investigation and even finding the right questions and approaches are challenging tasks. Unavoidably, complexity science should have a significant role in the endeavor of exploring this ``terra incognita”. Using its tools ranging from non-linear dynamics to network science and the theory of collective phenomena we have learned much about the human society~\cite{Ball2012Complexity}. Extending the results to the human-AI system and discovering the related new phenomena will be the goals of the period ahead.

Our main focus here stems from a paramount consideration of historical perspective. It is expected that the advent of machines that cover part of the intellectual activities of humans (AI agents) is as unavoidable as the advent of those machines that came to help our physical work during the industrial revolution. At that time, in fact, the production of work from energy started to be shifted from the human and animal body to that of combustion engines. The progressive lowering of costs of both energy sources and work production from those machines pushed toward an unavoidable increase in their number. The lowering costs of information processing and the parallel increase of their abilities up to the modern deep learning marvels is analogously pushing the increase of their number and frequency of use. A specific problem we are interested in is understanding how the Human-AI ecosystem reacts therefore to an increase of the relative fraction of machines $\alpha$. In particular, will the ecosystem react smoothly or abruptly? In the first scenario humans can optimize the outcomes by feedback while in the second there could be irreversible consequences.

Our plan is to describe the Human-AI ecosystem using the complex systems, statistical physics approach which has a well developed framework to spot abrupt changes and phase transitions.

Complex systems consist of interacting agents spanning a network. In social systems the nodes are individuals and the links between pairs of them stand for the different kinds of relationships humans may have. Recently the importance of so-called higher order interactions has been emphasized~\cite{ACMM2017, Battiston2020Beyond_pairwise,Bianconi2021Book}, where interactions beyond pairs are considered and processes on such networks have been studied, including opinion dynamics. From the mathematical point of view such generalization implies going from a simple graph-theoretical setting of vertices and edges to a more complex hyper-graph environment where also three-body terms {\it faces} or higher are considered. For example, in social systems the binding of a triple is significantly different from the sum of three links between its components. Analogously the presence of different types of agents, i.e. the nature of the agent inhabiting a node, represents a further element of complexity to be considered (see for instance \cite{McFadden2001}). 

In the present paper we focus on ecosystems with both humans and AI agents with the presence of binary and higher order interactions. There are several reasons why such a system cannot be discussed in its full complexity. There is no generally accepted model of the problem, and we are not aware of the details of the interactions. In such a situation there are two main and in some sense opposite routes of theoretical approaches. First, one can set up a complex agent-based model where the numerous parameters can be target of a complex fitting procedure~\cite{Murase2021DeepLearning} to mimic behaviour which is thought to be reasonable. Second, one can chose an oversimplified model, where some basic elements of the original problem are present and where the analytical approach can be pushed to its limits. We can expect from such a model insight into the qualitative behaviour of the system, information about possible phases and the transitions between them. We follow in this paper the second route. 

The first approach of this type can be traced back to Daniel McFadden's Discrete Choice theory~\cite{McFadden2001}. It is possible to show indeed that such theory is equivalent to a multi-populated model without interaction or, in other words, a model of independent agents belonging to a finite number of groups~\cite{GalloBarra2009}. In that model each agent aims to optimize its utility function up to some fluctuation of logistic type. In~\cite{Durlauf1996a,Durlauf1996b,BrockDurlauf2001} the model, for a single group case, was generalized to include a mean field interaction between agents. A full interacting generalization of the McFadden model to the multi-group case came in~\cite{ContucciGhirlanda2007, BarraContucci2014, OpokuOsabuteyK2019, BurioniContucci2015} where specific case studies were investigated. It is important to emphasize that, for all the mentioned examples, the Gibbs distribution is only a working hypothesis to be used as a possible guide and, at best, to be tested against data: a distribution is assumed and the free parameters it depends on are statistically inferred. For an information theoretical perspective see~\cite{Bialek2012, Jaynes1957, McKay2003}. 

As the starting point we have chosen the Ising model~\cite{Brush1982History}, the fundamental model of statistical physics, which was originally designed to describe the paramagnetic-ferromagnetic phase transition. It consists of agents with binary state variables represented by nodes of a network, interacting with their neighbors, where interactions prefer alike or opposite states. A transition probability of the state variables depending on the neighbors’ states and the noise level (temperature) completes the definition of the model.  
Later on we will define the model in mathematical terms. The Ising model has been solved analytically in the limit of large number of agents on the complete graph (mean-field solution) and on two-dimensional lattices. Here solution means the description of the system in its equilibrium, stationary state. It is worth noticing that Ising models with cubic interaction were studied in \cite{SubLebowitz99} in the context of occupation number variables (i.e. taking $0$ and $1$ values). In that study, the phase diagram and the segregation kinetics of a system of particles on a two-dimensional square lattice were investigated. The Ising model has been extensively used for modelling social phenomena~\cite{Galam2008}.

In this work we extend the original mean-field Ising model in two ways. First, we have introduced two kinds of agents for humans and AI units, both having binary state variables, but the interaction between the different types of agents can be different. It should be noted that such a generalization to a two-component model has been studied earlier in a different context~\cite{CGM08a, GC08}. Furthermore, we introduce higher order interactions to investigate the effects mentioned above. Our task is to calculate the phases and the transitions between them as a function of the coupling parameters and, especially, of the relative size of the two components that we call $\alpha\in [0,1]$. When $\alpha=1/2$ the two groups, for instance Human and AI would be of the same size, $\alpha=0$ would mean only Human agents are present and vice versa $\alpha=1$ only AI. Our aim is to gain insight into the possible behaviour of the human-AI ecosystem. 

The paper is organized as follows. In the next section, we define the mathematical model with three-body interaction. Section 3 is devoted to the exploration of the results. We describe the solution with one component, and solve the two-component model. The analysis of the phases, and the determination of the order of the transitions are also discussed in the same section.  General remarks and perspectives are discussed in the final section Conclusion and
Outlook.

\section{The model and mathematical results}
The human-AI ecosystem is an interacting system with two components, i.e.,  two kinds of entities. There are many ways to model mathematically such a system, depending on the characterization of the entities, the types of interactions, and the topology of the interactions. Our choice has been governed by the following viewpoints. We wanted to have an analytically tractable model, as such a solution enables general conclusions and qualitative predictions. Furthermore, we wanted to work with a model, which is closely related to standard theoretical models to be able to tie in with traditional techniques and results. Our choice has been 
a specific class of solvable Ising models that we can analytically control and make qualitative predictions with. Applications of the same class of models have been used in different fields such as health \cite{BurioniContucci2015}, immigration \cite{ContucciVernia2020,BarraContucci2014, GalloBarra2009}, education \cite{OpokuOsabuteyK2019}, energy conservation \cite{OsabuteyOpokuS2020}, and protein structure \cite{MorcosE1293}, among others. The paradigm we're considering therefore doesn't have to be understood as strictly 
typical to AI and Human agents; it is rather a flexible analytical guide to a variety of phenomena characterised by ecosystems of multi-components interacting agents.


The interactions that will be taken into account are the binary, quadratic ones (H-H, H-AI and AI-AI), the triplet or cubic (AI-AI-AI, AI-AI-H, AI-H-H, H-H-H) while the higher order ones (quartic, etc.) will be ignored. Furthermore, we consider the state (or opinion) variables of the agents as binary having in mind a task with possible binary outcome, like should a patient be operated or not. The nature of the interaction is that two or more agents in contact may like (or dislike) each others ``opinion" and they change their state variables such that the system finds a stationary state.
We assume that all agents are in contact with each other, i.e., they sit on the nodes of a complete graph - an assumption needed for the analytical solvability of the system.

The model described above can be formulated mathematically as follows. Let us consider a system of $N$ interacting agents, where  each agent $i$ has an internal degree of freedom described by a spin variable $\sigma_i \in \{-1,1\}$ that represents the agent's opinion. A configuration of the system is then determined by a vector $\sigma \in \Sigma_N=\{-1,1\}^N$. The class of models we are interested in is described by a \textit{Hamiltonian function} or \textit{cost function} of the following form  

\begin{equation}\label{3CCW}
    \mathcal{H}_N(\sigma) = - \sum_{i,j,k = 1}^N K_{i,j,k}\, \sigma_i\sigma_j\sigma_k - \sum_{i,j = 1}^N J_{i,j} \sigma_i\sigma_j - \sum_{i = 1}^N h_i\sigma_i \; ,
\end{equation}
where $K_{i,j,k},J_{i,j}$ are families of real parameters that tune the interactions among agents and the $h_i$ tunes the bias of each agent. Eq \eqref{3CCW} is the Hamiltonian of the Ising model with two-body and three-body interactions. Eq \eqref{3CCW} is very general and it naturally includes the possibility of describing a two-component system, like the Human-AI ecosystem.

The parameter $J_{i,j}$ tunes the interaction among the couple of agents, while the $K_{i,j,k}$ those among triples. The positivity of those parameters imply, for the GKS inequalities \cite{Griffiths1967, KellySherman1986}, that the configurations with aligned opinions are the favored ones. The collective properties of the equilibrium state of the system are codified by a Boltzmann-Gibbs type probability distribution, related to the cost function \eqref{3CCW}

\begin{equation} \label{Gibbs}
    \mu_{N} (\sigma) = \frac{e^{- \mathcal{H}_N(\sigma)}}{Z_N} 
\end{equation}
where the normalization factor, $Z_N$ also called the partition function, is used to compute the generating function of the moments for the previous distribution \eqref{Gibbs} 

\begin{equation}\label{FEnergy}
p_N\,=\, \dfrac{1}{N}\,\log Z_N  \; ,
\end{equation}
which coincides, up to a multiplicative factor, with the free energy of the model.

Clearly, the large $N$ behaviour or thermodynamic limit (TDL) of $p_N$ depends on the choice of the parameters of the Hamiltonian \eqref{3CCW}. In the sequel we will show how to represents the large number limit of $p_N$ as a finite dimensional variational problem in two specific mean-field cases.

\subsection{The one component cubic mean-field model} \label{CCW_LD}

In the sequel we consider only mean-field interaction among agents, i.e. we make the assumption based on the full permutation symmetry among agents, $K_{i,j,k}=K/3N^2$, $J_{i,j}=J/2N$ and $h_i=h$.
The resulting model turns out to be a cubic mean-field model:

\begin{equation}\label{1CCW}
\mathcal{H}_N(\sigma) = - N\left(\frac{K}{3} m_N^3(\sigma)+ \frac{J}{2} m_N^2(\sigma) + h m_N(\sigma)\right)
,
\end{equation}  
where $m_N(\sigma)=\frac{1}{N} \sum_{i = 1}^{N} \sigma_i$ is average opinion. In eq \eqref{1CCW} $K$ and $J$ are the cubic and binary couplings, respectively, and $h$ is the uniform bias. A remarkable fact about the considered model is that it may be solved exactly \cite{GO2023} by means of the {\it large deviations} technique, a method developed in \cite{Ellis85}.
One can prove that the large limit number of the generating functional \eqref{FEnergy} related to the Hamiltonian \eqref{1CCW} admits the following variational representation:
\begin{equation}\label{var1CCW}
p(K,J,h) := \lim_{N\to\infty} p_N = \sup_{m\in[-1,1]} \phi(m),
\end{equation}
where $\phi(m)= U(m)- I(m)$ with
\begin{equation}\label{energy}
U(m)= \dfrac{K}{3} m^3 +\dfrac{J}{2} m^2 +h m 
\end{equation}
is the energy contribution and
\begin{equation}\label{entropy}
I(m)=\frac{1-m}{2}\log\left(\frac{1-m}{2}\right)+\frac{1+m}{2}\log\left(\frac{1+m}{2}\right)
\end{equation}
is the entropy contribution, expressing the logarithm of the number ways the value $m$ can be produced with different $\sigma$ configurations. 


The structure of the probability measure identified by the variational principle \eqref{var1CCW} select stable solutions, i.e. a small stochastic disturbance of the system will produce small changes on the opinions unless the system is close to a second order phase transition.

The solutions of the variational principle \eqref{var1CCW} must satisfy the stationarity condition
\begin{equation}\label{mfeq1}
    \overline{m}=\tanh( K\overline{m}^2 +  J\overline{m} +   h),
\end{equation}
the equation can be solved \cite{GO2023} by means of the local fixed-point method.
Among those solutions we are only interested in the ones that realize the supremum of $\phi$ in \eqref{var1CCW} since they represent the overall opinion of the system with respect to the measure \eqref{Gibbs} in the equilibrium state. The quantity $\overline{m}$ is called the order parameter of the model. The overall phase picture that emerges presents novel features.

\begin{figure}[htp]
    \centering
    \includegraphics[width=8cm]{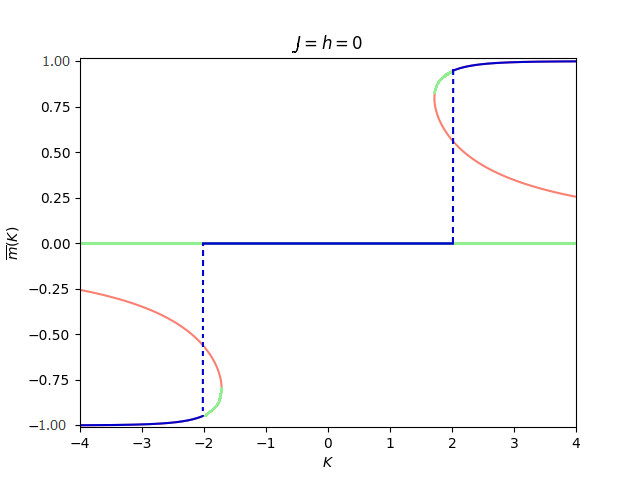}
    \caption{\small  \textbf{Average opinion, $\overline{m}$, of the system for $J=h=0$ as a function of $K$.} There is a transition in $\overline{m}$ from zero to positive average opinion when crossing $K = 2.016295$ from below and a transition from zero to negative average opinion when crossing $K=-2.016295$ from above.  The curves represent all the solutions of the stationary condition \eqref{mfeq1}, the blue ones corresponds to the global stable ones, i.e., the solution that realize the supremum of $\phi$, the green ones are locally stable solutions and the red ones are the unstable solutions.}
    \label{Fig1}
\end{figure}

Unlike the quadratic mean field model that, for $h=0$, has a second order continuous phase transition in $J$, the cubic case analyzed here displays a remarkable \textit{discontinuous first order phase transition}
in $K$ when $J=h=0$ shown in Fig \ref{Fig1}. Starting from small absolute valued $K$-s and increasing or decreasing it, the value $\overline{m}$ characterizing the stable stationary solution remains at zero until $K=K_c\approx  \pm 2.016295$ where suddenly we observe a jump in the order parameter. 

\begin{figure}[htp]
    \centering
    \includegraphics[width=16cm]{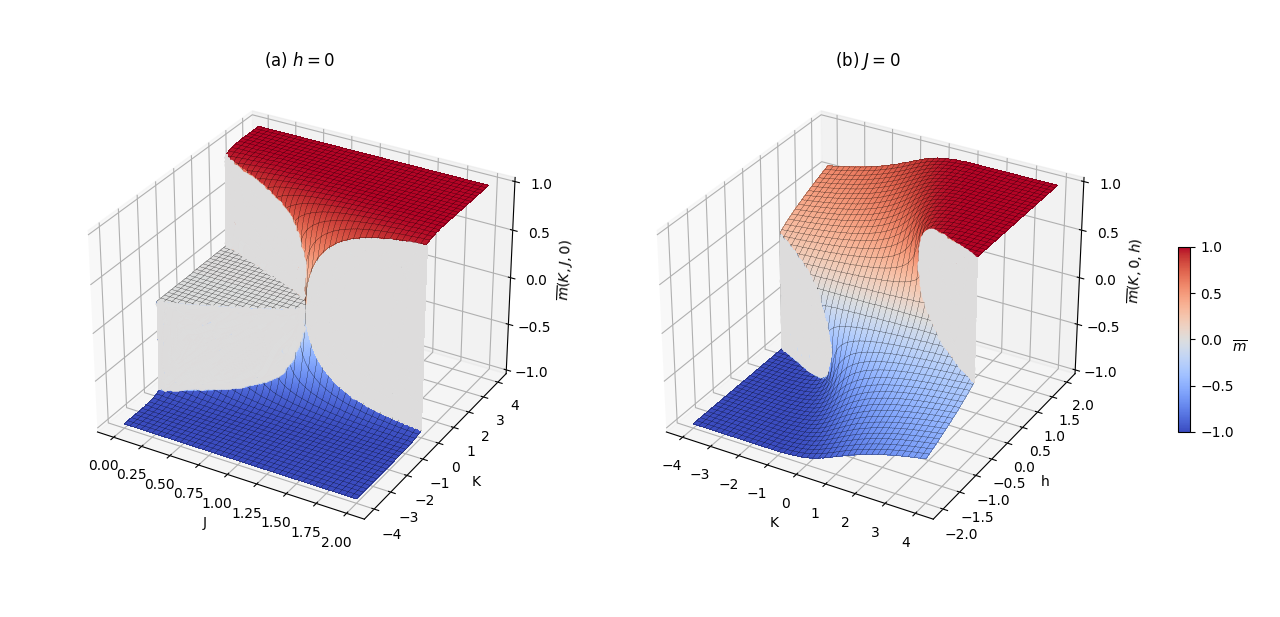}
    \caption{\small \textbf{Average opinion surfaces of the cubic mean-field model.} In panel \textbf{(a)}, $h=0$ while in \textbf{(b)} $J=0$. When $J=0$ in \textbf{(a)}, we observe the global stable solution found in Fig \ref{Fig1} which indicates jumps at $K_c$ and when $K=0$, we obtain the solution of the simple Ising model without cubic interaction. For fixed $J<1$ and moving along $K$ the system presents two jumps separated by a plateau at zero. Those two jumps coalesce into a single one when $J$ cross the unit value. In \textbf{(b)} we observe a discontinuity in the average opinion for two separated jumps when $h$ and $K$ falls within certain thresholds.}
    \label{Fig2}
\end{figure}

The behaviour of the order parameter in the planes $(K,J,0)$ and $(K,0,h)$ is shown in Fig \ref{Fig2}, while the case $(0,J,h)$ correspond to the classical two-body mean-field model. From panel (a) one can observe the presence, for $J<1$, of three distinct phases: the one with negative average opinion (in blue), the one with zero average opinion (in gray) and the one with positive average opinion (red). In that region therefore a progressive increase in $K$ from negative to positive values encounter two consecutive jumps.

\newpage
\subsection{The two component cubic mean-field model}
Our main interest is to understand the behaviour of a system consisting of two kinds of agents, corresponding to AI and H. To this end we investigate the two component version of the model in eq \eqref{3CCW}, defined as follows. Let partition the system of $N$ agents into two subsystems AI and H of sizes $N_1$ and $N_2$ respectively, such that AI $\cap$ H = $\emptyset $ and $N_1 + N_2 = N$. Let $m_S(\sigma)=\frac{1}{|S|} \sum_{i\in S } \sigma_i$ be the average opinion of agents in a subsystem $S$ and denote $m_1$ and $m_2$ as the average opinion for the subsystems AI and H respectively.  Further, we define as the relative sizes of AI and H agents $\alpha_1 = \frac{N_1}{N}$ and $\alpha_2=\frac{N_2}{N}$ respectively. The two component cubic mean-field model has the following energy contribution:

\begin{equation}\label{2B-energy}
\begin{split}
U(m_1,m_2)  &= \frac{1}{3}\left[K_{111}\alpha_1^3m_1^3 + 3K_{112}\alpha_1^2\alpha_2m_1^2m_2+3K_{122}\alpha_1\alpha_2^2m_1m_2^2 + K_{222}\alpha_2^3m^3_2\right]\cr
&+\frac{1}{2}\left[J_{11}\alpha_1^2m_{1}^2 + 2J_{12}\alpha_1\alpha_2m_{1}m_{2} +  J_{22}\alpha_2^2m^2_{2}\right] + \left[h_{1}\alpha_1m_{1} + h_{2}\alpha_2m_{2}\right].
\end{split}
\end{equation}

\begin{figure}
\includegraphics[width=16cm]{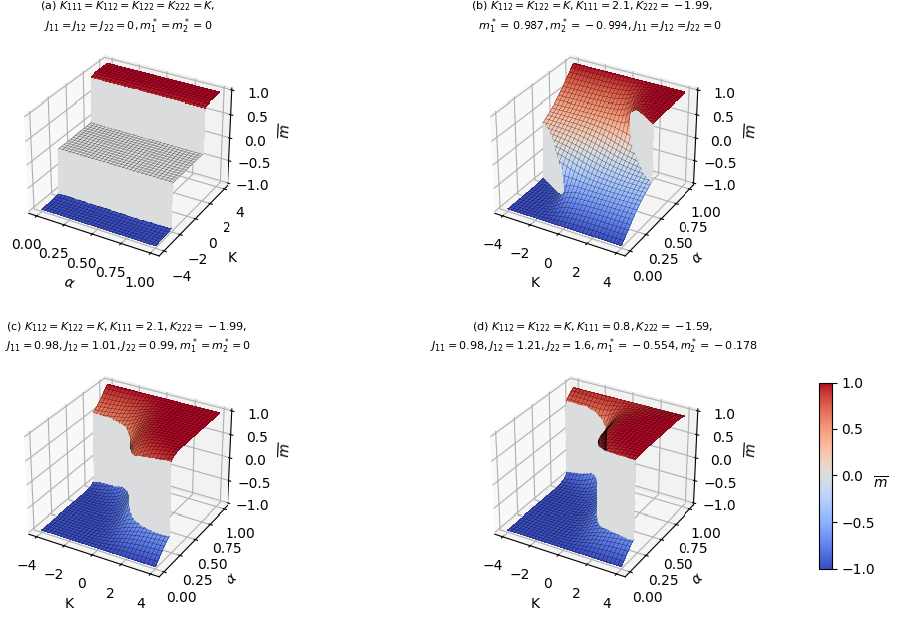}
\caption{\small \textbf{Total average opinion surfaces of the two component cubic mean-field model.} In panel \textbf{(a)} we observe first order phase transitions at $K_c$. Here, $\alpha$ is constant in $K$. When the cubic interactions are fixed (i.e.$K_{111}=K_{112}=K_{122}=K_{222}=K$) the proportion of AI and Human agents present in the system has no effect on their average opinion as observed in panel \textbf{(a)}. Two distinct jumps in $\overline{m}$ are observed in panel \textbf{(b)} for certain values of $K$ and $\alpha$. For panel \textbf{(c)} and \textbf{(d)}  $\alpha$ varies smoothly for the total average opinion and then observe sudden jump to another phase.}
\label{Fig3}
\end{figure}

The variational form of the large number limit of the generating functional \eqref{FEnergy} associated to the  two component cubic mean-field model \eqref{2B-energy} is \cite{GO2023}:
\begin{equation}\label{2B-VFE}
\sup_{m\in[-1,1]^2} \Phi(m) = \sup_{m\in[-1,1]^2}\left[U(m_1,m_2) - (\alpha_1 I(m_1)+ \alpha_2 I(m_2))\right]
\end{equation}
where $I(m_1)$ and $I(m_2)$ are the entropy associated to the average opinions of the subsystems and they sum up to the total number of configurations as a product of the individual ones. The stationary solutions $\overline{m}$ of $\Phi$ are as follows

\begin{equation}\label{2Bmag_CCW}
    \overline{m}_l = \tanh{\left(h_l + \sum_{p,q=1}^2\alpha_p(J_{lp}+\alpha_qK_{lpq}\overline{m}_q)\overline{m}_p \right)} \qquad \qquad \text{for} \qquad l=1,2, 
\end{equation}
from which the global stable ones are to be selected. In the rest of this work, we assign $\alpha = \alpha_1$ and $(1-\alpha)=\alpha_2$ then $\alpha \in [0,1]$ and the total average opinion $\overline{m}=\alpha\overline{m}_1+(1-\alpha)\overline{m}_2$ will be used as combined order parameter.
It is worth recalling that, when $\alpha=0$ then there are only Human agents in the population and when $\alpha=1$ there are only AI agents in the population. For the rest of the work, we adopt the re-parameterisation of $h_1$ and $h_2$ found in \cite{CGM08a}. In this sense, the parameters $h_1$ and $h_2$ are thought of as dependent on the internal average opinion (given by $m_1^*$ and $m_2^*$) and interaction within each subsystem without interaction with the other agents. Hence following the fixed point method, we define $h_1$ and $h_2$ as follows;

\begin{equation}
\begin{split}
    h_1 =& \arctanh(m_1^*) - K_{111}{m_1^*}^2 - J_{11}m_1^*\cr
    h_2 =& \arctanh(m_2^*) - K_{222}{m_2^*}^2 - J_{22}m_2^*
\end{split}
\end{equation}
Surfaces of the solution of \eqref{2Bmag_CCW} that gives rise to the global maxima of $\Phi$ in eq \eqref{2B-VFE}, with respect to the free parameters $\alpha$ and $K_{112}=K_{122}=K$ for fixed values of the other parameters are shown as Fig \ref{Fig3}. 

When cross cubic interactions (i.e. $K_{112}=K_{122}=K$) are fixed, as observed in panels \textbf{(b), (c)} and \textbf{(d)} of Fig \ref{Fig3}, there are jumps in the average opinion of the agents depending on their relative fractions. Smaller values of $\alpha$ (i.e. more Human agents), may lead to an inclination of the minus opinion while positive opinion inclination may result from larger values of $\alpha$ (i.e. more AI agents). Hence, a larger proportion of the AI agent population may lead to abrupt changes in behaviour of the ecosystem.

\begin{figure}
\includegraphics[width=18cm]{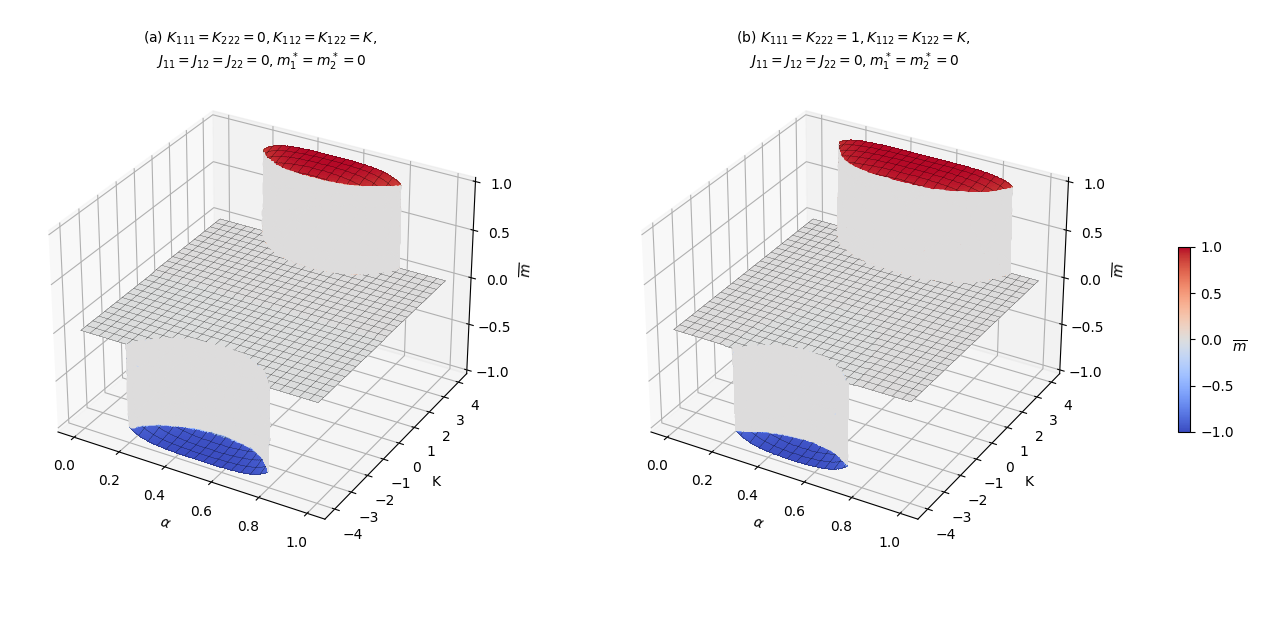}
\caption{\small \textbf{Average opinion for $K_{111}=K_{222}=0$ and $K_{111}=K_{222}=1$ with $K_{112}=K_{122}=K$ varying.} In the left panel, panel \textbf{(a)}, the cubic in-group interaction  for the AI agents and that of humans are set to zero (i.e. $K_{111}=K_{222}=0$) and in \textbf{(b)} to one (i.e. $K_{111}=K_{222}=1$) with varying inter-group interaction.}
\label{Fig4}
\end{figure}

AI  machines are made to assist and work with humans, therefore they are assumed to interact with humans. Fig \ref{Fig4} gives the scenario of an interacting system where only Humans or only AI agents are not interacting among themselves (see panel \textbf{(a)}) and when we assume that the cubic interaction among humans and AI agents are equal (see panel \textbf{(b)}). In both cases we observe transitions for large enough fraction of the AI agents in the order parameter when interaction bias and mutual interaction are absent.

\subsection{Exploration of the effect of the composition}
\begin{figure}[!h]
\includegraphics[width=16cm]{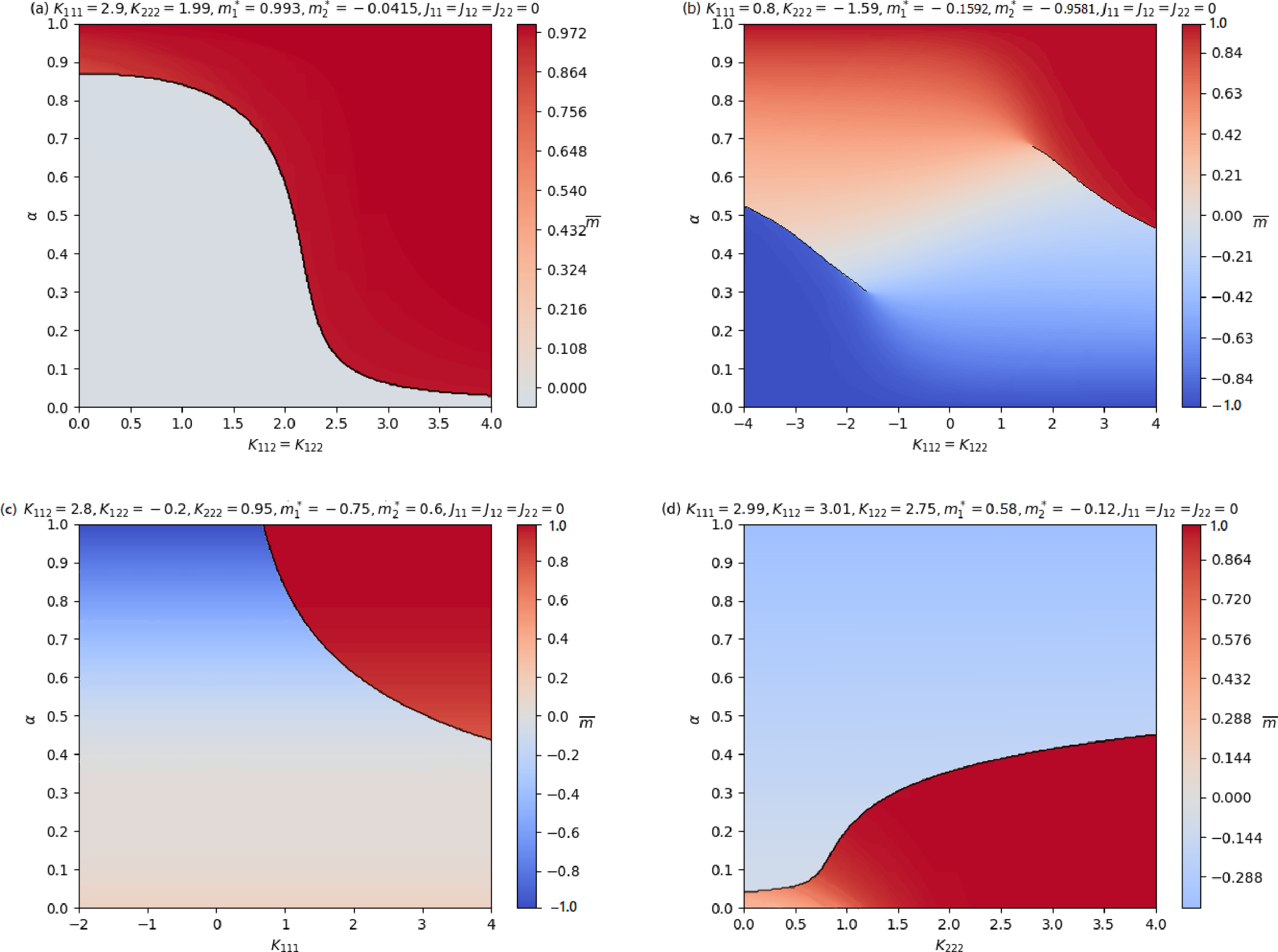}
\caption{\textbf{Phase diagram for fixed parameters of the cubic mean-field model.} In panel \textbf{(a)}, when $K_{112}=K_{122}$ is small the system require a larger fraction (i.e. $\alpha$) of the AI agents to observe a phase transition and as $K_{112}=K_{122}$ increases, the proportion of AI agents required for a phase transition decreases. In this scenario, the relative fraction of the AI agents corresponding to the black line is  a decreasing function of $K_{112}=K_{122}$. While panel \textbf{(b)} illustrate two separate jumps in the average opinion depending on the values of $K_{112}=K_{122}$ and $\alpha$. We observe from panel \textbf{(c)}  that when interaction among AI agents ($K_{111}$) increases their proportion needed to observe a phase transition decreases. While in panel \textbf{(d)}, when interaction among Human agents ($K_{222}$) increases, the fraction of AI agents needed to effect a phase transition increases and vice versa.
}
\label{Fig5}
\end{figure}

In this section we present phase diagrams of the model relating the parameter $\alpha$ and one among the interacting ones. The black continuous line is used to separate the opinion phases and in particular it emphasises the first order phase transitions, i.e. sudden jumps of the opinion resulting in abrupt changes of color in the picture.
This is illustrated in Fig \ref{Fig5}. The $\alpha$-value found in correspondence of the black line indicates the proportion of AI agents required for Human opinion to lose its prevalence over the entire population. 

The simulation of the results obtained for suitable values of the model parameter in Fig \ref{Fig5}\textbf{(a)} suggest that even in the case where there is very small fraction of AI agents we can still observe abrupt behaviours in opinion formation within the Human AI ecosystem. This observation is likewise similar to that of panel \textbf{(d)} of Fig \ref{Fig5}, which illustrate that for a system where there is less interaction among Human agents ($K_{222}$), smaller fraction of the AI agents may lead to phase transition and hence prevalent opinion formation.

\section{Conclusions and Outlooks}
The rapid spread of AI raises sharply the question of the relationship between humans and machine intelligence. How far can we control this development? Will we become vulnerable or will we succeed in humanising this new form of collaboration~\cite{XDGG2021TransHCI}? In fact, the problems are already all around us. What happens when AI and human participants in a medical consultation take different positions? Can trust or reticence develop between representatives of different origins? Can group interactions play a role? Our simple model aims to clarify such questions.

With the advancement of AI systems, there are growing concerns among scientists and psychologists about the creation of a machine world with AI and its eventual replacement of people in the labour force~\cite{HankMusk2015, RDT2015}. In light of this possible occurrence, the following works~\cite{LFBL2018, Hancock2019, SSACDDZ2019, Salmon2019, Shneiderman2020a, Shneiderman2020b} advise users and developers, as well as humans in general, to take an ethical approach to the design and use of these AI technologies, ensuring that humans are at the center of the algorithms with AI machines.

Notwithstanding these important factors, AI machines are needed to meet our daily needs as humans. To illustrate the significance of this study, let us first consider the undoubted benefit of AI machines to our healthcare system. Starting from chronic diseases and cancer to radiography and risk assessment, there are virtually endless opportunities to leverage technology to deliver more precise, efficient, and forceful interventions at precisely the appropriate moment in a patient's care. Given the massive amount of data accessible, artificial intelligence is clearly positioned to be the primary engine that will propel our healthcare system forward as need dictates. 

In this study we propose a simple, analytically solvable, mathematical model with higher order interaction that seeks to model the complex interactions between humans and AI agents. Using this model we predict and emphasize some crucial features of the system that are triggered on its composition i.e. the proportion of AI participating in a joint decision with us. This aspect is especially important in the light of the new tendency toward distributed AI, when the interaction of a large number of AI units is be more the standard than the exception~\cite{Eisenstadt2018}.
The model utilized here has the potential to forecast the likely outcome in opinion formation (such as whether a patient should be operated on or not) for a certain proportion of AI or Human agents as they interact. Abrupt changes, known as discontinuous phase transitions in statistical physics, may predict thresholds beyond which AI machines may gain control prevalence. Experts in human-computer interaction, industries, institutions, psychologists, and others who work with AI machines may rely on the use of these models to set boundaries in order to achieve the desired and ethical benefit to humans while reducing its negative impact. The simulations conducted with our model (see Fig \ref{Fig5}) show that a seemingly modest proportion of AI agents may lead to their prevalence in opinion formation over humans. 

A noteworthy feature of the cubic mean-field model stems from the fact that we can observe three distinct phases depending on the parameter values in the absence of interaction bias of the agent(s) in the ecosystem (see for instance Fig \ref{Fig2}(a), Fig \ref{Fig3}(a) and Fig \ref{Fig4}). 
Unlike the quadratic mean-field model, where we observe a jump from negative to positive state, instead one can observe a jump from the negative average opinion to a zero average opinion and a jump to a positive average opinion when three body interaction is considered. The zero average opinion, which is a stable paramagnetic state, is an indication of symmetry in opinion such that the agents have no preference for one over the other. As $K$ increases or decreases, the symmetry in opinion is broken, and the total average opinion of the agents in the ecosystem shifts to either a positive or negative state. 

The results illustrated in this work are some of the possible simulation for a wide class of values of the parameters. The model used and the whole statistical mechanics approach presented might also be used to infer the values of those parameters starting from real data as it was done in ~\cite{BurioniContucci2015, ContucciGhirlanda2007, BarraContucci2014, OpokuOsabuteyK2019}.

Clearly, our approach has limitations. First, the dynamics leading to the stationary statistical distribution described by eq \eqref{FEnergy} is a special one, while in reality opinion dynamics may be quite different as reflected in the numerous models introduced to study it~\cite{XiaXuan2011OpRev}. However, we believe that our simplified model is sufficient for calling the attention to a possible source of criticality, namely that the dependence of the outcome in an AI/H ecosystem depends very non-linearly on the composition of the participants and this may have severe consequences.

A further aspect is that in realistic settings time should play an important role, which has been completely ignored here. With reference to the example above about the decision making process in a critical situation in health care, there is probably not enough time to achieve a complete equilibrium state of the participating opinion carriers. Another source of non-stationarity could be that the system is driven by a continuous flow of data. Therefore, more realistic models have to be dynamic in nature.

The above critical points give a guide to us in which direction one should continue the research on the Human-AI ecosystem. An important step should be to collect and use data of related systems as a starting point for the developments of adequate models.


\section*{Acknowledgement}
The research reported in this work was partially supported by the EU H2020 ICT48 project „Humane AI Net“ under contract \# 952026, by the European Union – Horizon 2020 Program under the scheme “INFRAIA-01-2018-2019 – Integrating Activities for Advanced Communities”, Grant Agreement n.871042, “SoBigData++: European Integrated Infrastructure for Social Mining and Big Data Analytics”, and by the CHIST-ERA grant "SAI": CHIST-ERA-19-XAI-010, FWF (grant No. I 5205).


\begin{thebibliography}{}
\bibitem{Russel2021AI} Russell S, Norvig P. Artificial intelligence: A modern approach. 4th ed. Upper Saddle River, NJ: Pearson; 2020.

\bibitem{Ball2012Complexity} Ball P. Why society is a complex matter: Meeting twenty-first century challenges with a new kind of science. 2012th ed. Berlin, Germany: Springer; 2012.

\bibitem{ACMM2017} Alberici D, Contucci P, Mingione E, Molari M. Aggregation models on hypergraphs. Ann Phys (N Y). 2017;376:412–424. doi: 10.1016/j.aop.2016.12.001

\bibitem{Battiston2020Beyond_pairwise} Battiston F, Cencetti G, Iacopini I, Latora V, Lucas M, Patania A, Young J-G, Petri G. Networks beyond pairwise interactions: Structure and dynamics. Phys Rep. 2020;874:1–92. doi: 10.1016/j.physrep.2020.05.004.

\bibitem{Bianconi2021Book}
Bianconi G. Higher-Order Networks. Cambridge: Cambridge University Press; 2021. (Elements in Structure and Dynamics of Complex Networks).

\bibitem{McFadden2001} McFadden D. Economic choices. Am Econ Rev. 2001;91:351–378. doi: 10.1257/aer.91.3.351.

\bibitem{Murase2021DeepLearning}	Murase Y, Jo H-H, Török J, Kertész J, Kaski K. Deep learning based parameter search for an agent based social network model. arXiv [physics.soc-ph]. 2021. Available from: http://arxiv.org/abs/2107.06507.

\bibitem{GalloBarra2009}	Gallo I, Barra A, Contucci P. A minimal model for the imitative behaviour in social decision making: theory and comparison with real data. Math Models Methods Appl Sci. 2009;19.

\bibitem{Durlauf1996a} Durlauf SN. of NBER, technical working paper series. Statistical Mechanics Approach to Socioeconomic Behavior. 1996;203.

\bibitem{Durlauf1996b} Durlauf SN. How can statistical mechanics contribute to social science? Proc. Proc Natl Acad Sci USA. 1996;96:10582–10584.

\bibitem{BrockDurlauf2001} Brock WA, Durlauf SN. Discrete choice with social interactions Rev. Rev Econ Stud. 2001;68:235–260.

\bibitem{BurioniContucci2015} Burioni R, Contucci P, Fedele M, Vernia C, Vezzani A. Enhancing participation to health screening campaigns by group interactions. Sci Rep [Internet]. 2015;5:9904. doi: 10.1038/srep09904. Cited: in: : PMID: 25905450.

\bibitem{ContucciVernia2020} Contucci P, Vernia C. On a statistical mechanics approach to some problems of the social sciences. Front Phys. 2020;8. doi: 10.3389/fphy.2020.585383.

\bibitem{ContucciGhirlanda2007} Contucci P, Ghirlanda S. Modeling society with statistical mechanics: an application to cultural contact and immigration. Qual Quant. 2007;41:569–578. doi: 10.1007/s11135-007-9071-9.

\bibitem{BarraContucci2014} Barra A, Contucci P, Sandell R, Vernia C. An analysis of a large dataset on immigrant integration in Spain. The Statistical Mechanics perspective on Social Action. Sci Rep. 2015;4. doi: 10.1038/srep04174.

\bibitem{OpokuOsabuteyK2019}Opoku AA, Osabutey G, Kwofie C. Parameter evaluation for a statistical mechanical model for binary choice with social interaction. J Probab Stat. 2019;2019:1–10. doi: 10.1155/2019/3435626.


\bibitem{Bialek2012} Bialek W, Cavagna A, Giardina I, Mora T, Silvestri E, Viale M, Walczak AM. Statistical mechanics for natural flocks of birds. Proc Natl Acad Sci U S A. 2012;109:4786–4791. doi: 10.1073/pnas.1118633109. Cited: in: : PMID: 22427355.

\bibitem{Jaynes1957} Jaynes ET. Information theory and statistical mechanics. Phys Rev. 1957;106:620–630. doi: 10.1103/physrev.106.620.

\bibitem{McKay2003} Mckay D. Information Theory, Inference, and Learning Algorithms. Cambridge: Cambridge University Press; 2003.


\bibitem{Brush1982History} Brush SG. History of the Lenz-Ising model. Reviews of modern physics. 1967;39.

\bibitem{SubLebowitz99} Subramanian B, Lebowitz J. The study of a three-body interaction Hamiltonian on a lattice. J Phys A Math Gen. 1999;32:6239–6246. doi: 10.1088/0305-4470/32/35/302.

\bibitem{Galam2008}  Galam S. Sociophysics: A review of Galam models. Int J Mod Phys C. 2008;19:409–440. doi: 10.1142/s0129183108012297.

\bibitem{CGM08a} Contucci P, Gallo I, Menconi G. Phase transitions in social sciences: two-population mean field theory, Int. Int Jou Mod Phys B. 2008;22:1–14.

\bibitem{GC08} Gallo I, Contucci P. Bipartite Mean Field Spin Systems. Existence and Solution. Math Phys E J. 2008;14.

\bibitem{OsabuteyOpokuS2020} Osabutey G, Opoku AA, Gyamfi S. A statistical mechanics approach to the study of energy use behaviour. J Appl Math. 2020;2020:1–14. doi: 10.1155/2020/7384053.


\bibitem{MorcosE1293} Morcos F, Pagnani A, Lunt B, Bertolino A, Marks DS, Sander C, et al. Direct-coupling analysis of residue coevolution captures native contacts across many protein families. Proc Natl Acad Sci U S A. 2011;108:E1293-301. doi: 10.1073/pnas.1111471108. PMID: 22106262.

\bibitem{Griffiths1967} Griffiths RB. A Proof that the Free Energy of a Spin System is Extensive. J Math Phys. 1964;5:1215–1222. doi: 10.1063/1.1704228; Correlations in Ising Ferromagnets. I. J Math Phys. 1967;8:478–483. doi: 10.1063/1.1705219.

\bibitem{KellySherman1986}Kelly DG, Sherman S. General Griffiths’ Inequalities on Correlations in Ising Ferromagnets. J Math Phys. 1968;9:466–484. doi: 10.1063/1.1664600.

\bibitem{GO2023} Osabutey G. \textit{PhD Thesis in preparation}. Alma Mater Studiorum - University of Bologna, Italy. 

\bibitem{Ellis85} Ellis RS. Entropy, large deviations, and statistical mechanics. 2006th ed. Berlin, Germany: Springer; 2005.

\bibitem{XDGG2021TransHCI} Xu W, Dainoff MJ, Ge L, Gao Z. Transitioning to human interaction with AI systems: New challenges and opportunities for HCI professionals to enable human-centered AI. arXiv [cs.HC] [Preprint] 2021. Available from: http://arxiv.org/abs/2105.05424.

\bibitem{HankMusk2015} Hawking S, Musk E, Wozniak S. Autonomous weapons: an open letter from AI and robotics researchers. Future of Life Institute. 2015.

\bibitem{RDT2015} Russell S, Dewey D, Tegmark M. Research priorities for robust and beneficial artificial intelligence. AI Mag. 2015;36:105–114. doi: 10.1609/aimag.v36i4.2577.

\bibitem{LFBL2018} Lau N, Fridman L, Borghetti BJ, Lee JD. Machine learning and human factors: Status, applications, and future directions. Proc Hum Factors Ergon Soc Annu Meet. 2018;62:135–138. doi: 10.1177/1541931218621031.

\bibitem{Hancock2019} Hancock PA. Some pitfalls in the promises of automated and autonomous vehicles. Ergonomics. 2019;62:479–495. doi: 10.1080/00140139.2018.1498136.

\bibitem{SSACDDZ2019} Stephanidis C, Salvendy G, Antona M, Chen JYC, Dong J, Duffy VG, et al. Seven HCI grand challenges. Int J Hum Comput Interact. 2019;35:1229–1269. doi: 10.1080/10447318.2019.1619259.

\bibitem{Salmon2019} Salmon PM. The horse has bolted! Why human factors and ergonomics has to catch up with autonomous vehicles (and other advanced forms of automation): Commentary on Hancock (2019) Some pitfalls in the promises of automated and autonomous vehicles. Ergonomics. 2019;62:502–504. doi: 10.1080/00140139.2018.1563333.

\bibitem{Shneiderman2020a} Shneiderman B. Human-centered artificial intelligence: Reliable, safe and trustworthy. Int J Hum Comput Interact. 2020;36:495–504. doi: 10.1080/10447318.2020.1741118.

\bibitem{Shneiderman2020b} Shneiderman B. Design lessons from AI’s two grand goals: Human emulation and useful applications. IEEE Trans Technol Soc. 2020;1:73–82. doi: 10.1109/tts.2020.2992669.

\bibitem{Eisenstadt2018} Eisenstadt V, Espinoza-Stapelfeld C, Mikyas A, Althoff K-D. Explainable distributed case-based support systems: Patterns for enhancement and validation of design recommendations. Case-Based Reasoning Research and Development. Cham: Springer International Publishing; 2018. p. 78–94.

\bibitem{XiaXuan2011OpRev}
Xia H, Xuan Z. Opinion Dynamics: A Multidisciplinary Review and Perspective on Future Research. International Journal of Knowledge and Systems Science. 2011;2:72–91.
\end{thebibliography}
\end{document}